\documentclass[numberedappendix]{emulateapj}
\usepackage{natbib}
\slugcomment{ApJ in press}

\shorttitle{Strangulation in Galaxy Groups}
\shortauthors{Kawata and Mulchaey}

\begin{document}


\title{Strangulation in Galaxy Groups}

\author{Daisuke Kawata\altaffilmark{1,2}
and John S. Mulchaey\altaffilmark{1}
}

\altaffiltext{1}{The Observatories of the Carnegie Institution of Washington,
 813 Santa Barbara Street, Pasadena, CA 91101.
}
\altaffiltext{2}{
 Swinburne University of Technology, Hawthorn VIC 3122, Australia.
}

\begin{abstract}

 We use a cosmological chemodynamical simulation 
to study how the group environment impacts the star formation properties of 
disk galaxies. The simulated group has a
total mass of  M $\sim$ 8 $\times$ 10$^{12}$ M$_{\odot}$ and 
a total X-ray luminosity of
L$_{\rm X}$ $\sim$ 10$^{41}$ erg s$^{-1}$. 
Our simulation suggests that ram pressure is not sufficient in this group to remove
the cold disk gas from a V$_{\rm rot}$ $\sim$ 150 km s$^{-1}$ galaxy.
However, the majority of the hot gas in the galaxy is stripped over a timescale of 
approximately 1 Gyr.
Since the cooling of the hot gas component provides a source for new cold gas, the stripping of the 
hot component effectively cuts off the supply of cold gas. 
This in turn leads to a quenching of star formation.
%The galaxy maintains the disk component
%after the cold gas is consumed leading to a galaxy with S0 properties.
The galaxy maintains the disk component
after the cold gas is consumed, which may lead to a galaxy similar to an S0.
Our self-consistent simulation suggests that this strangulation 
mechanism works even in low mass groups,
providing an explanation for the lower star formation rates 
in group galaxies relative to galaxies in the field.
\end{abstract}

\section{Introduction}
\label{sec-intro}

Understanding the role environment plays in determining the star formation properties of 
galaxies
remains one of 
the most important issues in galaxy evolution. 
Since the work of \citet{hh31},
many studies have shown that clusters of galaxies contain a higher fraction of 
early-type galaxies
\citep[e.g.,][]{ao74,ad80} and fewer star forming galaxies
\citep[e.g.,][]{bmy97,gos03} 
than the field. Several physical mechanisms have been
proposed to explain this trend. The most promising mechanisms
fall into one of three broad categories:
(1) mergers and interactions with other galaxies 
\citep[e.g.,][]{eh41,tt72,sw78} and/or the cluster potential 
\citep[e.g.,][]{bv90,mkldo96,kb99};
(2) ram-pressure 
and/or viscous stripping of the cold gas 
\citep[e.g.,][]{gg72,pn82,rfs80,fn99,amb99,qmb00}; (3)
\lq\lq strangulation\rq\rq \ in which the warm and hot gas in the halo
is stripped cutting off the supply of cold gas
\citep[e.g.,][]{ltc80,bnm00,bcs02}.

Recent observations suggest that
star formation is also suppressed
in some groups of galaxies 
\citep[e.g.,][]{zm98a,wbb05b,wvym06}.
While galaxy interactions and mergers are often assumed to play 
a dominant role in driving galaxy evolution in galaxy groups 
\citep[e.g.,][]{hrt77,ccs81,imtks83,jb89}, the recent observations of
both X-ray and HI tails in nearby groups 
\citep[e.g.,][]{dkmh97,ssc04,mnsjf05,rpm06,bc02} suggests other mechanisms 
may also be important.
Unfortunately, most studies have concentrated on rich galaxy clusters, and 
much less is known about the mechanisms important
in lower mass groups \citep[e.g.,][]{yf04,rh05,jh06,mff07}.
 In this {\it Letter} we use a self-consistent 
cosmological chemodynamical simulation to demonstrate that
strangulation can effectively suppress star formation 
in low mass groups.
%We describe our simulation in Section \ref{sec-meth} and 
%examine Section \ref{sec-res}. 

%\input{tables}
%\input{figures}
\begin{figure*}
%\plotone{eps/evolbw_c.ps}
\plotone{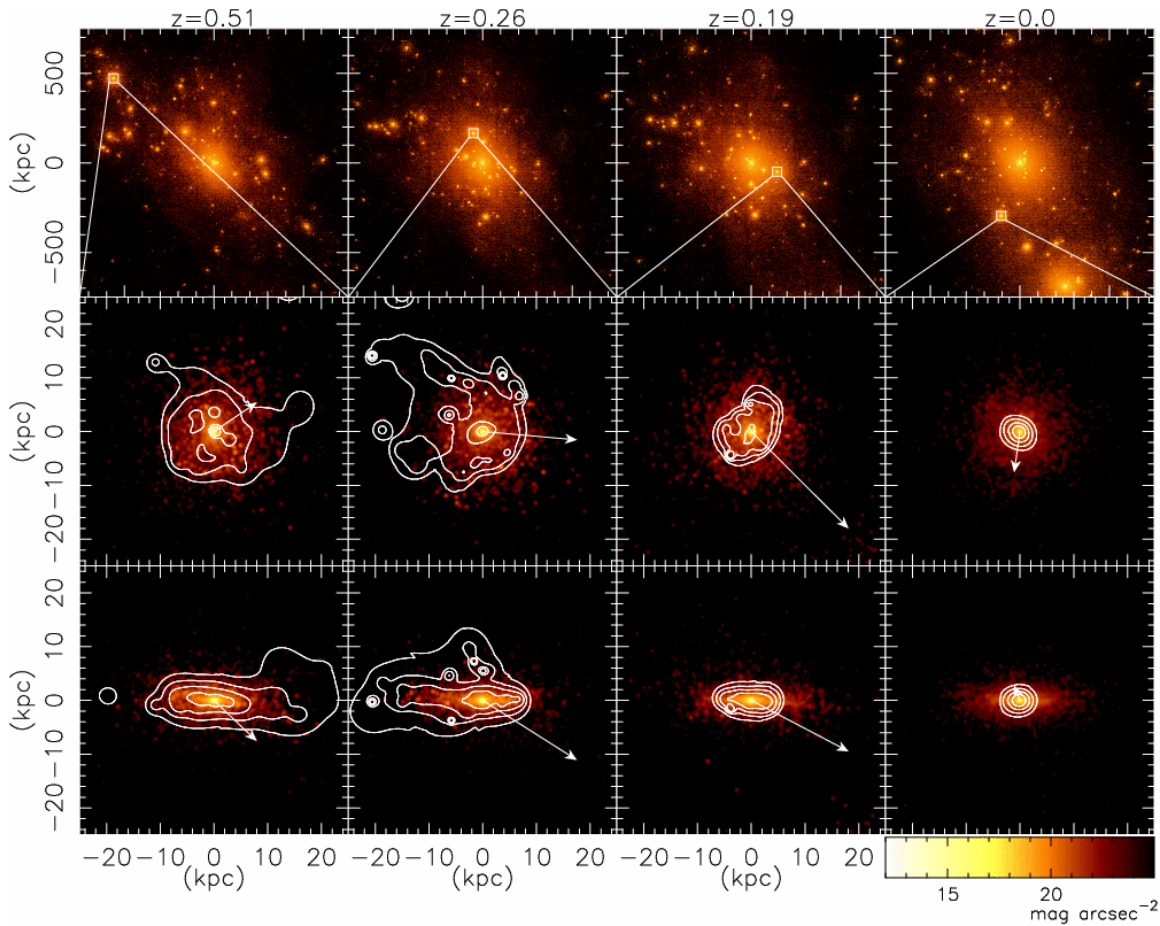}
\caption{
 Evolution of the distribution of dark matter density 
from an arbitrary but fixed projection
({\it top rows}), and the close-up rest-frame $R$-band images of the
target galaxy from the face-on ({\it middle rows}) and edge-on 
({\it bottom rows}) views. 
The contours correspond to the cold gas density
%of $\log (\rho_{\rm cold}/{\rm g\ cm}^{-2})=0.2$, 0.6, 1.2, 1.6.
of 
$\log (\rho_{\rm cold}/{\rm M_{\sun} pc}^{-2})=0.2, 0.6, 1.2, 1.6$.
The arrows indicate the velocity of the target galaxy
with respect to the velocity of the group.
The length of the arrows of 1 kpc
correspond to the speed of 20 km s$^{-1}$.
\citet{bc03} model is adopted for the stellar population synthesis.
\label{fig-evol}}
\end{figure*}

\begin{figure}
%\plotone{eps/sforb.ps}
\epsscale{0.7}
\plotone{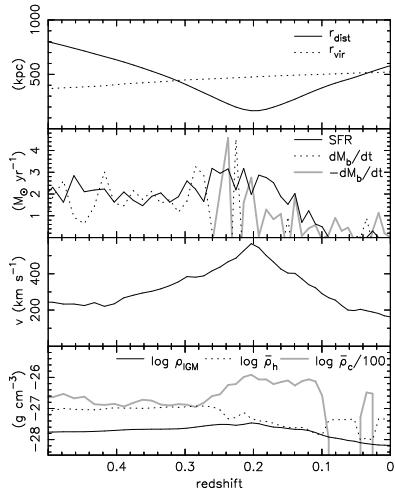}
\caption{
{\it Top}: The virial radius of the group ({\it dotted})
and the distance of the target galaxy from the group center ({\it solid})
as a function of redshift. 
{\it Second}: The star formation rate  ({\it black solid}),
the net mass-accretion ({\it dotted}), and mass-loss ({\it gray solid}) rates 
of baryon within 30 kpc of the target galaxy as a function of redshift.
{\it Third}: Relative speed of the target galaxy to the group
as a function of redshift. 
{\it Bottom}: The intra-group gas density
at the position of the target galaxy ({\it solid}) and the hot ({\it dotted})
and cold ({\it gray solid}) gas densities of the target galaxy
as a function of redshift.
We define $\rho_{\rm IGM}$ as the volume weighted mean density of
the hot gas between the radius of 50 and 100 kpc from the target galaxy.
The hot (cold) gas density indicates the volume (mass) weighted
mean density within 30 kpc. Note that the cold gas density is
scaled by $1/100$.
%({\it fourth}) and ram pressure ({\it bottom})
% as a function of redshift.
%The unit of ram-pressure is 
%$10^{10}$~M$_{\sun}$~km$^{2}$~s$^{-2}$~kpc$^{-3}$.
%
% z=0.3: 10.402 Gyr
% z=0.2: 11.368 Gyr
% z=0.1: 12.467 Gyr
% z=0  : 13.719 Gyr
\label{fig-sforb}}
\end{figure}

\begin{figure}
%\plotone{eps/rf.ps}
\plotone{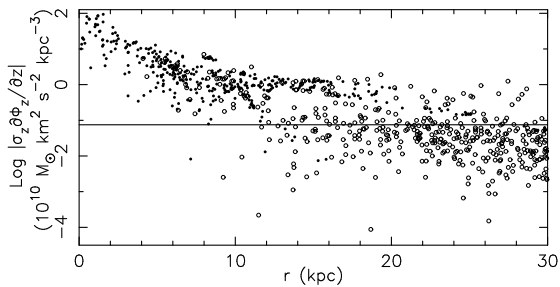}
\caption{
 The estimated restoring force, 
$\sigma_z |\partial \Phi_z/\partial z|$, for the cold (filled circles)
and hot (open circles) gas particles as a function of their radius
at $z=0.26$, where $\sigma_z$ and $\partial \Phi_z/\partial z$ 
are respectively the surface gas density and the restoring 
gravitational acceleration along $z$ direction
which is set to the direction of the relative velocity of the target 
galaxy to the group.
The horizontal line shows the estimated ram-pressure 
at the position of the target galaxy at $z=0.26$.
Note that the radius is the three-dimensional radius, and
the restoring force is estimated at the position of each particle,
i.e., we do not assume any symmetry, but use directly the particle
distribution. Therefore, particles with the same radius can have
different restoring force.
The cold gas particles at large radii ($r>10$ kpc) correspond to
the tail of the cold gas seen in Figure \ref{fig-evol}.
\label{fig-rf}}
\end{figure}

\begin{figure}
%\plotone{eps/fmhz026.ps}
\plotone{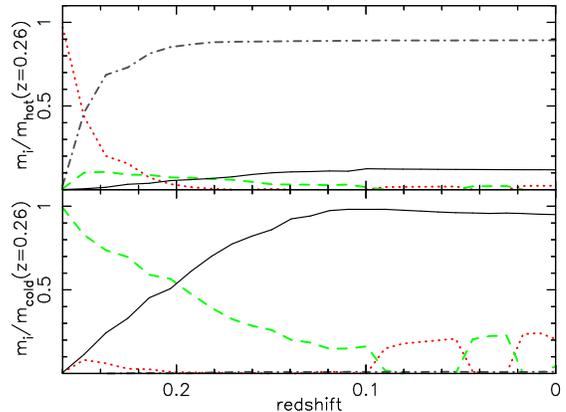}
\caption{
The history of the mass fraction of the hot gas ({\it dotted}), 
cold gas ({\it dashed}),
stars ({\it solid}), and the escaped ({\it dot dashed})
components which are originally from the hot
({\it upper}) and cold gas ({\it lower})
particles within 30 kpc of the target galaxy at $z=0.26$.
The mass fractions for the {\it upper} and {\it lower} panels
are normalized by the mass of the hot and cold gas at $z=0.26$.
The particles which are further than 30 kpc from the center
of the target galaxy are defined as escaped.
Note that the cold gas particles can become hot gas due to heating
by SN feedback. The mass of stellar particles decreases and
the gas particle mass can increase because of the mass-loss from stars.
\label{fig-fmh}}
\end{figure}

%\begin{deluxetable*}{ccccccccccccc}
\begin{deluxetable*}{lllllllllll}
%\begin{deluxetable}{lllllllllll}
\tabletypesize{\scriptsize}
\tablecolumns{11}
\tablewidth{0pc}
\tablecaption{General properties of the group at $z=0$ and 
 the target galaxy at $z=0.51$\label{tab-gprop}}
\tablehead{
 \multicolumn{5}{c}{the group at $z=0$} &
 \colhead{} &
 \multicolumn{5}{c}{the target galaxy at $z=0.51$} \\
 \cline{1-5} \cline{7-11} \\
 \colhead{$M_{\rm vir}$\tablenotemark{a}} & 
% \colhead{$R_{\rm vir}$\tablenotemark{a}} &
 \colhead{$M_{\rm gas,vir}$} & 
 \colhead{$M_{\rm DM,vir}$} & 
 \colhead{$M_{\rm star,vir}$} & 
 \colhead{$L_X$\tablenotemark{b}} & 
% \colhead{$T_X$\tablenotemark{b}} &
 \colhead{} &
 \colhead{$M_{\rm vir}$\tablenotemark{a}} & 
% \colhead{$R_{\rm vir}$\tablenotemark{a}} &
 \colhead{$M_{\rm gas,vir}$} & 
 \colhead{$M_{\rm DM,vir}$} & 
 \colhead{$M_{\rm star,vir}$} & 
 \colhead{$V_{\rm rot}$\tablenotemark{c}} \\
 \colhead{($M_{\sun}$)} &
% \colhead{(kpc)} & 
 \colhead{($M_{\sun}$)} & \colhead{($M_{\sun}$)} &  \colhead{($M_{\sun}$)} & 
 \colhead{(erg s$^{-1}$)} &
% \colhead{(keV)} & 
 \colhead{} &
 \colhead{($M_{\sun}$)} &
% \colhead{(kpc)} & 
 \colhead{($M_{\sun}$)} & \colhead{($M_{\sun}$)} &  \colhead{($M_{\sun}$)} & 
 \colhead{(km s$^{-1}$)}
}
\startdata
$8.1\times10^{12}$ &
% 522 &
$6.7\times10^{11}$ & $6.9\times10^{12}$ & $5.6\times10^{11}$ &
$8.76\times10^{40}$ & &
% L25n96/ran1/lev1/r4g2/sne3nth01/grbd/z000 group 1
%
$3.3\times10^{11}$ &
% 138 &
$2.3\times10^{10}$ & $2.7\times10^{11}$ & $3.4\times10^{10}$ &
150
% L25n96/ran1/lev1/r4g2/sne3nth01/grbd/z051 group 7, but not used
% L25n96/ran1/lev1/r4g2/sne3nth01/grbdb01n5e2/z051 group 21
\enddata
\tablenotetext{a}{The definition of \citet{ks96}}
\tablenotetext{b}{The X-ray luminosity in energy range between 0.01-100 keV
from the extrapolation of the modeled flux from the X-ray spectrum 
within r$_{500}=244$ kpc, where $r_{500}$ is the radius
of a sphere containing a mean density of 500 times the critical
density of $\rho_{\rm crit}=3 H_0/8\pi G$ at $z=0$.}
% L25n96/ran1/lev1/r4g2/sne3nth01/grbd/z000g1r10/x-ray/r0-r500xz
% flux x 4 pi (17 Mpc)^2 (1Mpc=3.086e24 cm) -> erg/s
% flux = 2.5324E-12 ergs cm**-2 s**-1
\tablenotetext{c}{The peak gas rotation velocity around the radius of 7 kpc.}
\end{deluxetable*}
%\end{deluxetable}

\section{Method}
\label{sec-meth}

The simulations were carried out using our original galactic chemodynamics
code, {\tt GCD+} \citep{kg03a,kg03b,kacg06,kr07}.
{\tt GCD+} is a three-dimensional tree $N$-body/smoothed
particle hydrodynamics \citep[SPH][]{ll77,gm77} 
code \citep{bh86,hk89,kwh96}
that incorporates self-gravity,
hydrodynamics, radiative cooling, star formation, supernova (SN)
feedback, and metal enrichment. {\tt GCD+} takes into account chemical
enrichment by both Type~II \citep{ww95}
and Type~Ia \citep{ibn99,ktn00} SN and mass loss
from intermediate-mass stars \citep{vdhg97}, 
and follows the chemical enrichment history
of both the stellar and gas components of the system.
The code also includes non-equilibrium chemical reactions of hydrogen and
helium species (H, H$^{+}$, He, He$^{+}$, He$^{++}$, H$_{2}$,
H$_{2}^{+}$, H$^{-}$) and their cooling processes,
following the method of \citet{aazn97}, \citet{azan97}, and \citet{gp98}.
We use the UV background spectrum suggested by \citet{hm01},
and take into account radiative cooling and heating due to heavy elements
\citep{rs77,ckor95}.

The system which we focus on here is one of the simulated groups of galaxies
in our ``Virtual Group Catalog'' (VGRC) project. 
The VGRC is an on-going numerical simulation campaign with a goal
of simulating more than 30 groups with a range of masses and 
formation histories at high enough resolution to resolve
the structures of the member galaxies. 
%The simulations are carried on supercomputers in
%Japan (VPP-5000 at CfCA/NAOJ, SX-6 at JSS/JAXA) and 
%Australia (Beowulf clusters at Swinburne and Altix 3700 Bx2 at APAC).
All groups in the VGRC are identified 
in cosmological simulations with a $\Lambda$CDM cosmology 
($\Omega_0$=0.24, $\Lambda_0$=0.76,
$\Omega_{\rm b}$=0.042, $h=0.73$, $\sigma_8=0.74$, and $n_s=0.95$)
consistent with the measured parameters from three-year 
{\it Wilkinson Microwave Anisotropy Probe} data
\citep{sbd07}. We use a multi-resolution technique to achieve high-resolution
in the regions of the identified groups, including the tidal forces from
neighboring large-scale structures.
The initial conditions for the simulations are constructed
using the public software {\tt LINGER} and {\tt GRAFIC2} 
\citep{eb01}. Gas dynamics and star formation are included only 
within the relevant high-resolution region,
the surrounding low-resolution region contributes to the high-resolution 
region only through gravity.
For this particular group simulation, the size of the high-resolution
region is about $\sim$4.5~Mpc at $z=0$, and the low-resolution region
is a sphere with diameter of about 35~Mpc. 
Consequently, the initial conditions consist of a total of 
1,234,583 dark matter (DM) particles and 783,424 gas particles.
The mass and softening lengths of individual gas (DM)
particles in the high-resolution region are $4.41\times10^6$
($2.08\times10^7$) M$_{\sun}$ and 0.96 (1.61) kpc, respectively.
The high-resolution region is chosen from a low-resolution simulation
as the region within 4 times the virial
radius of the identified group.
The general properties of the group are summarized in 
Table~\ref{tab-gprop}.
The simulation starts at $z=29.7$, and the initial temperature and
the fractions of hydrogen and helium species are calculated
by {\tt RECFAST} \citep{sss99,sss00}.
We turn on the UV background radiation at $z=6$ \citep[e.g.,][]{fnl01}.
Modeling of SN feedback is the most uncertain part in numerical
simulations of galaxy formation. In the VGRC project,
we have chosen a relatively strong feedback model,
where each SN yields the thermal energy of $3\times10^{51}$ erg.

\section{Results and Discussion}
\label{sec-res}

The results of the group simulation are summarized in 
Figures ~\ref{fig-evol} and ~\ref{fig-sforb}.
The top panel of Figure~\ref{fig-evol} shows the evolution
of the dark matter density of the simulated group since $z=0.51$. 
To study how the star formation properties of galaxies are 
impacted by the group, we follow the evolution of a disk galaxy
on its first passage through the group.
The properties of this galaxy (hereafter referred to as the target galaxy) before
it enters the group are summarized in
Table~\ref{tab-gprop}. The
 middle and bottom panels in Figure~\ref{fig-evol} show
face-on and edge-on views of the target galaxy, respectively.
The contours give the surface density of the cold gas mass
and the arrows show the velocity of the target galaxy
with respect to the group velocity.
Throughout the {\it Letter}, we define the cold (hot) gas
as the gas with $\log T<4.3$ ($\log T>4.3$).

The top panel of Figure~\ref{fig-sforb} displays the
time evolution of the distance of the target galaxy 
from the center of the group.
From this panel it can be seen that the target galaxy first falls into the group (i.e. 
within the virial radius) around $z=0.31$ and makes its closest approach to the 
group center at z $\sim$ 0.2 before moving beyond the
virial radius again by $z=0$. 
The second panel in Figure~\ref{fig-sforb} shows 
the star formation rate (SFR) in the target galaxy. Particularly noteworthy is the 
sharp drop in the SFR after the galaxy passes the group 
pericenter, with star formation effectively ceasing by
$z\sim0.1$. 
The sharp drop in the SFR after the galaxy enters the group suggests
the group environment can effectively quench star formation
even in relatively low mass groups.

After the galaxy enters the group, the 
cold gas displays a clear tail
morphology in the direction opposite of the galaxy's motion
(see the $z=0.26$ panel in Fig.~\ref{fig-evol}).
Such cold gas tails are often cited
as evidence for
ram-pressure stripping.
To examine this possibility in our simulated group, 
we track the relative speed, $v$, the hot gas density 
of the intra-group medium (IGM), $\rho_{\rm IGM}$,
at the position of the target galaxy
as a function of redshift {(Fig.~\ref{fig-sforb})}, 
from which the ram-pressure, P$_{\rm ram}=\rho_{\rm IGM} v^{2}$,
can be calculated.
In Figure~\ref{fig-rf}, we compare the estimated ram-pressure at $z=0.26$ with
the restoring force for the cold and hot gas particles within the
target galaxy.  
Although the effect of ram-pressure, which should be
time and spatial dependent, is more complicated than in the simulation, the
figure provides a qualitative comparison and 
supports our results.
Figure~\ref{fig-rf} demonstrates 
%($\log P_{\rm ram}\sim7.5\times10^{8}$M$_{\sun}$~km$^{2}$~s$^{-2}$~kpc$^{-3}$)
that the ram-pressure experienced by the target galaxy is well-below the
amount required to strip the cold gas. We also examine how the
fraction of the hot gas and cold gas particles within 30 kpc (an
arbitrary choice to include the majority of the cold gas but avoid
contamination from the IGM) of the target galaxy evolves from $z=0.26$
to $z=0.0$.  Figure~\ref{fig-fmh} shows that virtually none of the cold gas
escapes. Interestingly, even the particles in the cold gas tail at
$z=0.26$ do not escape from the galaxy. Thus, we conclude that
ram-pressure stripping of the cold gas component is not the cause of
the downturn in the SFR.

 However, the majority ($\sim90$ \%) of the hot gas is
indeed stripped since $z=0.26$.
In addition, a significant fraction of the hot gas is converted into cold gas.
This suggests that
hot gas halos are potentially an important source of 
cold gas in disk galaxies. Furthermore, since ram-pressure is strong
enough to strip the hot gas (Fig.~\ref{fig-rf}),
the new supply of cold gas is cut-off.
By $z\sim0.1$ almost all the cold gas has turned into stars and
star formation ceases.
This is also demonstrated in the second panel of Figure~\ref{fig-sforb} which
shows the {\it net} baryon mass-accretion and mass-loss rates of the target
galaxy, i.e., simply the difference in the total baryon mass divided
by the time between the simulation outputs. Before the target galaxy
falls into the group, the mass-accretion rate is comparable to the
SFR. However, the mass-loss due to the hot gas stripping becomes
significant around $z\sim0.26$, and the mass-accretion rate is no
longer sufficient to sustain the SFR.
Therefore, we conclude that 
star formation is quenched in the target galaxy because
of ``strangulation'' \citep{ltc80}.

 Figure~\ref{fig-sforb} shows some enhancement of star formation 
since $z\sim0.26$. This suggests that the group environment somehow 
enhances star formation,
which accelerates the strangulation process. The bottom panel of
Figure~\ref{fig-sforb} displays the hot and cold gas density in the target
galaxy. The hot gas density decreases since $z\sim0.26$ due to the
stripping and becomes close to the IGM density. On the other hand,
the density of the cold gas where stars form significantly increases.
One possibility is that the cold gas is compressed due to
ram-pressure, which leads to enhanced star formation 
\citep[see also][]{ss01,kbwc03}.

 Recently, \citet{ams07} have shown that the SPH scheme can
underestimate fluid instabilities, such as the Kelvin-Helmholtz
instability (KHI), which may be important for the stripping of cold
gas \citep{pn82,mb00,mkmw07}.
To test the impact of the KHI in our simulation, we follow \citet{mb00}
to estimate the timescale of the KHI (i.e., $\tau_{\rm KH}
=M_{\rm gas}(<r)/(\pi r^2 \rho_{\rm IGM} v)$) 
for the target galaxy at $z=0.26$ when the hot gas
stripping becomes significant. At $r=10$ kpc (an arbitrary choice close 
to a radius where the hot gas stripping is expected in Fig.~\ref{fig-rf}), 
we find that $\tau_ {\rm KH}\sim2.8$ Gyr 
($M_{\rm gas}(<r)=1.6\times10^9$ M$_{\sun}$, 
$\rho_{\rm IGM}=2.84\times10^{-28}$
g cm$^{-3}$, and $v=416$ km s$^{-1}$ at $z=0.26$).
This is longer
than the timescale of the strangulation process for the target galaxy,
i.e., between $z=0.26$ and 0.1 ($\sim1.7$ Gyr). Hence, the KHI is not the
dominant mechanism of the gas stripping for the target galaxy
\citep[see also][]{mmwsm06,mff07}.
Note that the effect of KHI is more complicated than
our simple estimate. Therefore, our simulation may underestimate
the stripping mass.
%
% z=0.26: 10.774 Gyr
% z=0.1: 12.467 Gyr

 It is worth noting that at $z=0$ the target
galaxy is once again beyond the virial radius of the group,
and still has no star formation. 
Although the target galaxy has consumed its cold gas disk
via star formation by $z=0$, the disk structure of the stellar component is 
still intact (Fig.~\ref{fig-evol}).
Therefore, the properties of the target galaxy
are morphologically similar to those of an S0 galaxy.
Thus, we suspect that 
strangulation may produce a populations of S0 galaxies
\citep{ltc80} in and around poor galaxy groups. 
The current simulation suggests it would be interesting 
to more quantitatively compare a statistical sample
of simulated galaxies with observed S0 galaxies in the future.

Recent observations of the spiral galaxy NGC~2276 in the
NGC~2300 group support our conclusion that strangulation is important in galaxy groups.
%NGC~2276 displays a prominent tail in both HI and X-ray emission {\citep{dkmh97,rpm06}.
\citet{rpm06} have performed a detailed study of this system with 
{\it Chandra} and conclude that ram-pressure stripping is not 
sufficient to strip the cold gas in the disk. However, NGC~2276 is losing gas from the hot phase
at a rate of $\sim$ 5 M$_{\odot}$ yr$^{-1}$. Given this mass loss rate, the galaxy will likely exhaust its
current supply of cold gas within 1--2 Gyr. Similar to what we find for our target galaxy, 
\citet{rpm06} conclude that once star formation has ceased, 
NGC~2276 will have properties similar to an S0 galaxy. 
The NGC~2300 group is more massive than the group in our present study. Our 
results suggest that evidence for strangulation may be found in lower 
mass systems.

The simulated group is a fairly relaxed system with
a dominant central elliptical galaxy.  
Strangulation is likely to be most effective in dynamically evolved groups because of their enhanced 
IGM. Furthermore, galaxies entering relaxed groups
will be subjected to stripping over larger path lengths
than galaxies entering dynamically young systems
which consist of individual halos and not a common envelope.
The presence of a significant X-ray emitting IGM is a good indicator that a group is 
relaxed \citep{jm00}.  As nearly all X-ray groups contain a dominant central elliptical galaxy, 
while systems dominated by spirals tend to be non-detected in X-rays 
\citep{mdmb03,op04},
strangulation is likely more effective in 
systems with a central dominant elliptical galaxy.
If true, this might provide an explanation for the recent finding by 
\citet{wvym06} that the
early-type fraction is higher in systems 
with an early-type central galaxy than in halos with a late-type
central galaxy (``galactic conformity'').
The larger sample of the simulated groups in the VGRC
will allow us to test this idea.

Although this {\it Letter} highlights only one case
of strangulation, we are seeing a similar suppression of
star formation in other groups 
in the VGRC. With the large range of properties among the 
VGRC, we hope to 
be able to quantify the importance of strangulation and 
determine how the suppression of star formation depends on
factors such as the galaxy's mass and orbit. The simulated groups in the 
VGRC will also allow us to explore the importance of other mechanisms 
such as galaxy interaction in driving galaxy evolution in groups.

\acknowledgments

%We acknowledge
%the Center for Computational Astrophysics, CfCA, of the National Astronomical
%Observatory, Japan (VPP5000 was used), the
%Institute of Space and Astronautical Science 
%of Japan Aerospace Exploration Agency, and
%the Australian Partnerships for Advanced
%Computing, where the numerical computations for this paper were
%performed. JSM acknowledges support from NASA grant NNG 04-536.
%This research is undertaken as part of the Commonwealth Cosmology
%Initiative (CCI: http://www.tecci.org), and international collaboration
%supported by the Australian Research Council.

We acknowledge CfCA(VPP5000)/NAOJ, SSS/JAXA
and APAC where the numerical computations for this paper were
performed. JSM acknowledges support from NASA grant NNG 04-536.
This research was supported in part by the KITP NSF grant, PHY05-51164
and undertaken as part of the CCI supported by the ARC 

%\bibliography{./dkref}

\end{document}